\documentclass[aps,pra,twocolumn,amsmath,amssymb,nofootinbib,showpacs,superscriptaddress]{revtex4-1}
\usepackage[english]{babel}
\usepackage{latexsym}
\usepackage{graphics}
\usepackage{graphicx}
\usepackage{epsfig}
\usepackage{color}
\usepackage{bm}
\usepackage{amsmath}
\usepackage{amssymb}
\usepackage{amsthm}
\usepackage{dcolumn}
\usepackage{bm}
\usepackage{float}
\usepackage{hyperref}
\usepackage{color}
\usepackage{epstopdf}
\usepackage{cleveref}
\usepackage[svgnames]{xcolor}
\usepackage{enumerate}
\usepackage{braket}
\hypersetup{hidelinks,colorlinks=true,allcolors=DarkBlue}

\usepackage{cancel}
\usepackage{tikz}
\hypersetup{hidelinks,colorlinks=true,allcolors=DarkBlue}

\theoremstyle{remark}

\begin{document}

\preprint{APS/123-QED}

\title{Quantum soft filtering for the improved security analysis \\ of the coherent one-way QKD protocol}

\author{D.A. Kronberg} 
\affiliation{Russian Quantum Center, Skolkovo, Moscow 143025, Russia}
\affiliation{Moscow Institute of Physics and Technology, Dolgoprudny, Moscow Region 141700, Russia} 
\affiliation{Department of Mathematical Methods for Quantum Technologies, Steklov Mathematical Institute of Russian Academy of Sciences, Moscow 119991, Russia}

\author{A.S. Nikolaeva}
\affiliation{Russian Quantum Center, Skolkovo, Moscow 143025, Russia}
\affiliation{Moscow Institute of Physics and Technology, Dolgoprudny, Moscow Region 141700, Russia} 

\author{Y.V. Kurochkin}
\affiliation{Russian Quantum Center, Skolkovo, Moscow 143025, Russia}
\affiliation{QRate, Skolkovo, Moscow 143025, Russia}

\author{A.K. Fedorov}\email{akf@rqc.ru}
\affiliation{Russian Quantum Center, Skolkovo, Moscow 143025, Russia}
\affiliation{Moscow Institute of Physics and Technology, Dolgoprudny, Moscow Region 141700, Russia} 

\date{\today}
\begin{abstract} 
A precise security analysis of practical quantum key distribution (QKD) systems is an important step for improving their performance. 
Here we consider a class of quantum soft filtering operations, which generalizes the unambiguous state discrimination (USD) technique.
These operations can be applied as a basis for a security analysis of the original coherent one-way (COW) QKD protocol since their application interpolates between beam-splitting (BS) and USD attacks.
We demonstrate that a zero-error attack based on quantum soft filtering operations gives a larger amount of the information for Eve at a given level of losses.
We calculate the Eve information as a function of the channel length. 
The efficiency of the proposed attack highly depends on the level of the monitoring under the maintenance of the statistics of control (decoy) states, and best-case results are achieved in the case of the absence of maintenance of control state statistics. 
Our results form additional requirements for the analysis of practical QKD systems based on the COW QKD protocol and its variants by providing an upper bound on the security.
\end{abstract}
\maketitle

\section{Introduction}

QKD is a method that allows establishing unconditionally secure communications between distinct legitimate parties~\cite{Gisin2002,Scarani2009,Lo2014,Lo2016}. 
An important step in the practical implementation of QKD devices is a comprehensive analysis both from the viewpoint of hardware attacks and subtle questions of realizing QKD protocols~\cite{Scarani2009}.
A peculiarity of the security analysis of QKD protocols is the fact that the eavesdropper is limited by laws of physics only, whereas all advanced technological and computational resources are available for attacks.
The significant progress on the security analysis of QKD protocols against general attacks has been performed~\cite{Lo1999,Shor2000,Biham2006,Mayers2001,Renner2005,RennerGisin2005}. 
Experimental and industrial realizations of QKD devices have also reached key distribution of reasonable rates over distances of 100 km (for a review, see Ref.~\cite{Lo2016}).  

The class of QKD protocols that is of specific interest for the analysis is related to methods, which are used in existing commercial QKD devices.
Examples include, in particular, the distributed-phase-reference (DPR) approach~\cite{Yamamoto2003,Yamamoto20032,Yamamoto2005,Yamamoto2006,Yamamoto2007,Gisin2004,Stucki2005,Stucki2009,Stucki20092}, specifically, 
differential-phase-shift~\cite{Yamamoto2003,Yamamoto20032,Yamamoto2005,Yamamoto2006,Yamamoto2007} and COW~\cite{Gisin2004,Stucki2005,Stucki2009,Stucki20092} QKD protocols.
A general security proof of DPR-QKD in a realistic setting has been missing, and only particular cases have been considered~\cite{Yamamoto20062,Branciard2007,Branciard2008,Zhao2008,Curty2007,Tsurumaru2007,Curty2008,Curty2009,Kronberg2017}. 
This problem has been recently particularly resolved, and lower bounds for a variant of the COW protocol have been obtained~\cite{Moroder2012}.
Recently, this issue has been studied in the context of the discrete-variable prepare-and-measure quantum networks~\cite{Lim2019},
where the security of modified versions of the COW protocol is considered as one of the applications of the developed toolbox.
The results of Ref.~\cite{Lim2019} improves the bound of Ref~\cite{Moroder2012}.
Nevertheless, the search for security bounds for various implementations of COW-like schemes still is an important task in the theory of quantum key distribution.
An important concept in the security analysis of COW-like protocols is zero-error attacks based on the USD technique~\cite{Branciard2007}.
In this case, Eve can take advantage only of the losses, so the attack combines three USD strategies and preserves all the observed detection rates in Bob's detectors. 

In this work, we consider a general class of quantum operations, so-called quantum soft filtering, which can be used for obtaining security bounds for COW-like protocols without the use of phase randomized states.
We demonstrate that in the context of the COW security analysis, quantum soft filtering operations interpolate between standard BS attack and USD technique.
If applied as a basis for an attack scenario in the COW QKD protocol, quantum soft filtering operations give a larger amount of the information for Eve at a given level of losses.
We calculate the Eve information as a function of the channel length and show the efficiency of attacks based on soft filtering operations highly depends on the level of the control under the maintenance of statistics of control (decoy) states.
Our results are of interest for obtaining upper bounds on the security for practical QKD systems based on the COW QKD protocol and its variants.
Although we stress the case of the absence of the phase randomization technique and grouping the signals into blocks (as it is suggested in Ref.~\cite{Moroder2012}), 
these methods are a potential solution to the attack scenario considered below.
This is due to the fact that soft filtering operations should be applied individually to states, so the structure of the operations should be changed for the case of the presence of phase randomization.
We note that the methods that are used in the modified protocols require the changes in the optical scheme of the protocol in comparison with the standard COW protocol, such as the use of photon number resolving detectors~\cite{Moroder2012}.

Our work is organized as follows.
In Sec.~\ref{sec:COW}, we briefly review the COW QKD protocol and zero-error attacks.
In Sec.~\ref{sec:soft}, we discuss quantum soft filtering operations.
In Sec.~\ref{sec:attack}, we obtain improved security bounds for the COW protocol using an attack based on quantum soft filtering operations.
We discuss the correspondence between quantum soft filtering and BS-based and USD-based attacks.
In Sec.~\ref{sec:results}. we discuss the obtained results and compare them with BS and USD attacks.  
We discuss main results and conclude in Sec.~\ref{sec:conclusion}.

\section{COW protocol and zero-error attacks}\label{sec:COW}

Here we review the idea of the COW protocol~\cite{Gisin2004,Stucki2005,Stucki2009,Stucki20092}. 
The basic scenario is that Alice encodes bits using time slots containing either vacuum $|0\rangle$ or signal $|\alpha\rangle$ states, where the mean number of photons $\mu=|\alpha|^2<1$.
The logical bit corresponds to sequences $|0\rangle|\alpha\rangle$ and $|\alpha\rangle|0\rangle$.
In order to prevent attacks, Alice also sends control states of the form $|\alpha\rangle|\alpha\rangle$ (they also called decoy states).
Alice produces each bit value with probability $(1-f)/2$, whereas with the probability $f$ she sends the decoy sequence. 

The prepared quantum states are then sent to Bob through a quantum channel with length $L$ and the corresponding transmittance $t=10^{-\delta L}$, where $\delta$ is the attenuation coefficient.
On the detection side, 
Bob registers the time-of-arrival of the photons on detectors $D_B$ for providing the raw key and checks the statistics for the detections on detectors $D_{M_i}$ 
in the monitoring line for destructive interference of decoy and information sequences~\cite{Stucki2009}.
After the formation of the raw key, Alice and Bob implement a post-processing procedure and, in particular, check the statistical parameters of the protocol.
Specifically, the important stage is to check detection rates.

That is why the most interesting class of attacks that should be taken into account for obtaining the corresponding bounds for COW-like QKD protocols is zero-error attacks~\cite{Branciard2007}.
This class of attacks does not introduce any errors in usually checked parameters of the protocol, such as quantum bit error rate (QBER) and the visibility of an interferometer.
The simplest example of such an attack is the BS attack.
In the BS attack, Eve simulates the lossy channel by extracting the $(1-t)$ fraction of the signal with a beam-splitter, and then sends expected fraction $t$ to Bob on a lossless line.
The main advantage of this attack is the fact that beam-splitter is equivalent to losses in the transmission channel, so the attack is always possible and is impossible to detect by monitoring the data of Alice and Bob.
However, this attack is not very powerful since the Eve information tends to the Holevo information~\cite{Holevo1973} of initial states rather than to unity. 

Another basis for zero-error attacks is to use USD technique.
In this attack, Eve tries to extract the full information from transmitted quantum states and block them once she fails.
The USD-based attack has been considered in detail in Refs.~\cite{Curty2007,Branciard2007}, where each of three USD strategies is to distinguish sequences that begin and end with the vacuum state, 
in particular sequences $\ket{0\alpha0}$, $\ket{0\alpha:\alpha0}$ and $\ket{0:\alpha\alpha:0}$, which allows leaving zero value of QBER and the visibility equal to unity.

\section{Quantum soft filtering}\label{sec:soft}

Here we consider quantum soft filtering operations, which as we demonstrate generalize the USD-technique.
These operations correspond to probabilistic information extraction.
For a pair of non-orthogonal states $|a_0\rangle$ and $|a_1\rangle$, quantum soft filtering operations in the Stinespring representation (also known as the Stinespring dilation) are as follows:
\begin{equation}\label{eq:soft}
	\begin{split}
	|a_0\rangle \rightarrow \sqrt{p}|b_0\rangle\otimes|s\rangle + \sqrt{1 - p}|0\rangle\otimes|f\rangle, \\
	|a_1\rangle \rightarrow \sqrt{p}|b_1\rangle\otimes|s\rangle + \sqrt{1 - p}|0\rangle\otimes|f\rangle,
	\end{split}
	\end{equation}
where $\langle b_0|b_1\rangle < \langle a_0|a_1\rangle$ and $\langle s|f\rangle = 0$. 
Eq.~(\ref{eq:soft}) has the following physical meaning:
One can either make the states $\{|a_0\rangle, |a_1\rangle\}$ more distinguishable with success probability $p$, 
that yields the ``success'' signal $|s\rangle$, or one can fail with the probability $1-p$ and get the failure signal $|f\rangle$. 

We note that auxiliary states $|s\rangle$ and $|f\rangle$ help to reveal whether the soft filtering operation was successful, or the states passed into vacuum. 
Due to the fact that $|s\rangle$ and $|f\rangle$ are orthogonal, one can distinguish successful and unsuccessful applications of the operations.

Unitarity of quantum soft filtering~(\ref{eq:soft}) gives the following condition on the conservation of the scalar product between the original and final states:
\begin{equation}
	\langle{a_0|a_1}\rangle=p\langle{b_0|b_1}\rangle-(1-p).
\end{equation}
Thus, from the unitary condition, the success probability is as follows:
\begin{equation}
	p = \frac{1 - \langle a_0|a_1\rangle}{1 - \langle b_0|b_1\rangle}.
\end{equation}
There are two notable particular cases of soft filtering operations.
First is ``doing nothing'', where $a_i$ and $b_i$ are the same and the success probability is equal to unity. 
Second is USD discrimination, where $\langle{b_0|b_1}\rangle=0$ and the success probability is $1-\langle{a_0|a_1}\rangle$.

We should note that the possibility to use filtering operations and their generalizations has been considered in various contexts, in particular for the analysis of other QKD protocols, such as BB84~\cite{Curty2005} and DPS~\cite{Curty2007,Kronberg2014}.

\section{Security bounds for COW using quantum soft filtering}\label{sec:attack}

Here we demonstrate that quantum soft filtering is helpful for obtaining security bounds for COW-like protocols in its original version.
The attack based on quantum soft filtering operations has a number of important features.
First, our attack is adaptive, i.e. the strategy of the attack depends on the results of the previous steps.
Second, in this attack Eve does not waste its resources, i.e. the possibility to block some states, on the discrimination between vacuum and decoy states.
This allows focusing the attention on the discrimination between information states since the position of the states will be announced later. 
Third, the attack is designed in a zero-error manner.
In the view of the fact that COW-like protocols are widely used in experiments on QKD, it represents an interesting scheme for illustrating the application of quantum soft filtering.

\subsection{Attack scenario}

The attack scenario assumes that Eve works with the tuples of the states, which begin and end with vacuum state $\ket{0}$ in analogy to the USD-based attack. 
Then some of these tuples should be blocked by Eve, and the rest reaches Bob with higher intensity so that the average detection rate remains the same as expected by Bob and the visibility remains constant. 
Eve should also preserve the percentage of control states so that the legitimate users could not detect the attack by the statistical changes. 
The attack consists of several stages.

(i) On the first stage, Eve tries to find a vacuum state by measuring each state while blocking the other signals where vacuum states were not obtained.
For an information state, the probability of successful vacuum discrimination ($Z$) is $1 - e^{-\mu_A}$.
	
(ii) On the second stage, Eve starts forming the tuple for the found position of the vacuum state.
Then she performs a small number $T_{SF1}$ of SF1 operations, which can be defined by the following Stinespring representation:
\begin{equation}\label{soft_filtering_for_COW1}
	\begin{split}
	&|\alpha\rangle|0\rangle \rightarrow \sqrt{p_1}|\beta\rangle|0\rangle\otimes|\varepsilon_1\rangle|0\rangle\otimes|s\rangle + \\
	&\sqrt{1 - p_1}|0\rangle|0\rangle\otimes|0\rangle|0\rangle\otimes|f\rangle, \\
	&|0\rangle|\alpha\rangle \rightarrow \sqrt{p_1}|0\rangle|\beta\rangle\otimes|0\rangle|\varepsilon_1\rangle\otimes|s\rangle + \\
	& \sqrt{1 - p_1}|0\rangle|0\rangle\otimes|0\rangle|0\rangle\otimes|f\rangle, \\
	&|\alpha\rangle|\alpha\rangle \rightarrow \sqrt{q_1}|\beta\rangle|\beta\rangle\otimes|E\rangle\otimes|s\rangle + \\
	& \sqrt{1 - q_1}|0\rangle|0\rangle\otimes|0\rangle|0\rangle\otimes|f\rangle,
	\end{split}
\end{equation}
where the states $\{|\beta\rangle|0\rangle, |0\rangle|\beta\rangle, |\beta\rangle|\beta\rangle\}$ belong to the Bob subspace, 
and the states $\{|s\rangle, |f\rangle\}$ and $\{|\varepsilon_1\rangle|0\rangle, |0\rangle|\varepsilon_1\rangle, |E\rangle\}$ belong to the Eve subspace. 
Here the state $|E\rangle$ belongs to the two-dimensional subspace spanned by the states $|\varepsilon_1\rangle|0\rangle$ and $|0\rangle|\varepsilon_1\rangle$, 
and if we denote $\cos\varepsilon_1' = e^{-|\varepsilon_1|^2}$, then 
\begin{equation}
	\langle E|\varepsilon_1\rangle|0\rangle = \langle E|0\rangle|\varepsilon_1\rangle = \cos\frac{\varepsilon_1' }{2}, 
\end{equation}
where $|E\rangle$ lies between $|\varepsilon_1\rangle|0\rangle$, and $|0\rangle|\varepsilon_1\rangle$.

In our attack strategy, Eve blocks the whole tuple in case any of these SF1 operations fail.

The SF1 operation is useful for increasing the portion of control states after the attack.
If we denote $\mu_A = |\alpha|^2$, $\mu_B = |\beta|^2$ and $\mu_{E1} = |\varepsilon_1|^2$, then unitary conditions read
\begin{equation}
\begin{split}
	p_1 &= \frac{1 - e^{-\mu_A}}{1 - e^{-(\mu_B + \mu_{E1})}},
\end{split}
\end{equation}
\begin{equation}
\begin{split}
e^{-\frac 12\mu_A} &= \sqrt{p_1 q_1}e^{-\frac 12\mu_B}\sqrt{\frac 12 (1 + e^{-\mu_{E1}})} + \\ 
& \sqrt{(1 - p_1)(1 - q_1)}.
\end{split}
\end{equation}

From this expression it is possible to extract the success probability $q_1$ for the control states as follows:
\begin{equation}
\begin{split}
	q_1=\cos^2\left[\arccos\left(\frac{\sqrt{p_1}\sqrt{\frac{1+e^{-\mu_{E1}}}{2}}e^{-\frac{\mu_B}{2}}}{p_1e^{-\mu_B}\left(\frac{1+e^{-\mu_{E1}}}{2}\right)+1-p_1}\right)\right.-\\
	-\left.\arccos\left(\frac{e^{-\mu_A}}{p_1e^{-\mu_B}\left(\frac{1+e^{-\mu_{E1}}}{2}\right)+1-p_1}\right)\right].
\end{split}
\end{equation}
One can see that if $\mu_B$ is not too large, then $q_1$ is higher than $p_1$. 
So in this case SF1 operation increases the portion of control states.
We note that for large values of $\mu_{E1}$ and not very large values of $\mu_B$, SF1 operation implements a sort of USD for the information states, but without modifying control states that reach the Bob side and without trying to discriminate between information and control states. 
This information will be obtained later.

(iii) On the third stage, Eve performs a number of SF2 operations, which are defined as follows:
\begin{equation}
	\begin{split}
	&|\alpha\rangle\!|0\rangle\!\rightarrow\! \sqrt{p_2}|\beta\rangle\!|0\rangle\!\otimes\!|\varepsilon_2\!\rangle\!|0\rangle\!\otimes\!|s\rangle\!+ \\ 
	&\!\sqrt{1\!-\!p_2}|0\rangle\!|0\rangle\!\otimes\!|0\rangle\!|0\rangle\!\otimes\!|f\rangle, \\
	&|0\rangle|\alpha\rangle\! \rightarrow \sqrt{p_2}|0\rangle\!|\beta\rangle\!\otimes\!|0\rangle\!|\varepsilon_2\rangle\!\otimes\!|s\rangle\!+ \\ 
	&\!\sqrt{1\!-\!p_2}|0\rangle\!|0\rangle\!\otimes\!|0\rangle\!|0\rangle\!\otimes\!|f\rangle, \\
	&|\alpha\rangle\!|\alpha\rangle\!\rightarrow \sqrt{q_2}|\beta\rangle\!|\beta\rangle\!\otimes\!|\varepsilon_2\rangle\!|\varepsilon_2\rangle\!\otimes\!|s\rangle\!+\\ 
	&\!\sqrt{1\!-\!q_2}|0\rangle\!|0\rangle\!\otimes\!|0\rangle\!|0\rangle\!\otimes\!|f\rangle,
	\end{split}
	\end{equation}
with similar notations $\mu_A = |\alpha|^2$, $\mu_B = |\beta|^2$ and $\mu_{E2} = |\varepsilon_2|^2$. 
The unitary conditions take a simple form
	\begin{equation}
	\begin{split}
	p_2 = \frac{1 - e^{-\mu_A}}{1 - e^{-(\mu_B + \mu_{E2})}}, \quad q_2 = p_2 e^{\mu_A - \mu_B - \mu_{E2}}.
	\end{split}
\end{equation}

SF2 operation applies to the sequence from left to right until the first failure or $T_{SF2}$ successes in a row.
Then Eve performs the vacuum discrimination operation over both Bob's and Eve's modes, from right to left starting from the latest signal with successful SF2 operation.
The probability of successful vacuum discrimination for any information state is $1 - e^{-(\mu_B + \mu_{E2})}$. 
We note that SF2 decreases the percentage of control states, which is helpful for the case when the following operation is the vacuum states $|0\rangle$ discrimination. 

After vacuum discrimination, the tuple is successfully formed and the attack is completed.

(iv) If Eve fails to find the vacuum in any positions with successful SF2 operation then attack is considered to be failed for this tuple and all the pulses are blocked. 

We note that the efficiency of the attack depends on how strictly it is required to maintain the statistics of control states in the pulses that reached Bob. 
If not to follow the statistics, no SF1 operations are needed. 
Then the attack becomes much more efficient, otherwise the attack is less efficient, but still should be taken into account at long distances.

Let us consider Eve's information in the case of successful attack and calculate it after the announcement of the positions of information states. 
If SF1 has been applied successfully for information states then the Eve information is given by the Holevo value of the states $|0\rangle|\varepsilon_1\rangle$ and $|\varepsilon_1\rangle|0\rangle$ in her quantum memory, 
which is equal to $\chi_1{=}h_2\left[\left(1{-}e^{-\mu_{E_1}}\right)/2\right]$.
The same holds for SF2 and the corresponding Eve information is as follows: $\chi_2{=}h_2\left[\left(1{-}e^{-\mu_{E_2}}\right)/2\right]$.

For the vacuum discrimination at the beginning and the end of the attack Eve obtains the full information or nothing with the probability equal to $1/2$ for both cases. 

\subsection{Example}

\begin{figure}
	\includegraphics[width=\linewidth]{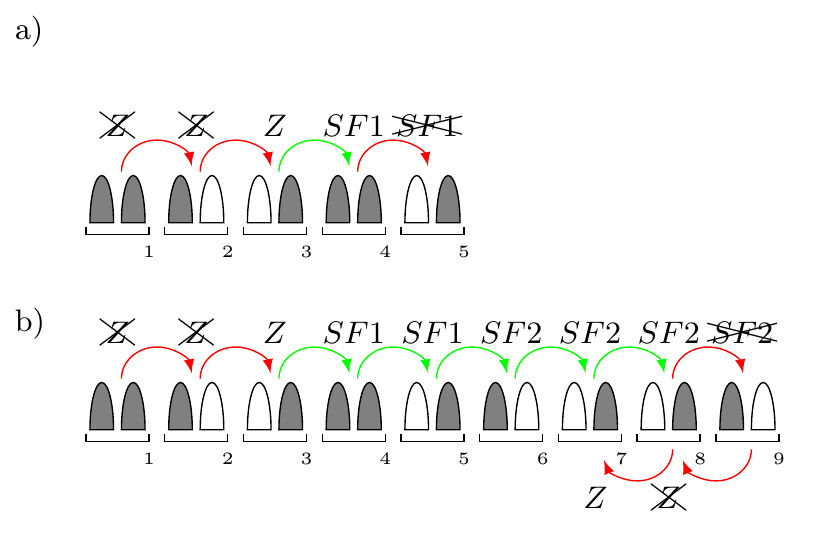}
\caption{Examples of the attack scenario.}
\label{fig:example}
\end{figure}

Let us consider two simple examples of the application of the attack. 
Suppose that the sequence of logical bits $c01c10110$ is as follows: 
\begin{equation}
\begin{aligned}
	|\alpha\rangle_1|\alpha\rangle_1|\alpha\rangle_2|0\rangle_2|0\rangle_3|\alpha\rangle_3|\alpha\rangle_4|\alpha\rangle_4|0\rangle_5|\alpha\rangle_5\\
	|\alpha\rangle_6|0\rangle_6|0\rangle_7|\alpha\rangle_7 0\rangle_8|\alpha\rangle_8|\alpha\rangle_9|0\rangle_9.
\end{aligned}
\end{equation}
where subindex stands for the signal number.

In the first case [see Fig.~\ref{fig:example}(a)] Eve is unable to find the vacuum states in first two signals (1 and 2) and she succeeds only in the case of third signal. 
Then for signal 4 Eve successfully realizes SF1 and for signal 5 SF1 fails. 
In this situation Eve does not get any information and should block all the signals from 1 to 5.

In the second case [see Fig.~\ref{fig:example}(b)] Eve is also able to find the vacuum states only in third signal. 
Then for signals 4 and 5 she successfully realizes SF1 operation and SF2 for 6, 7, and 8. In the case of signal 9 Eve does not succeed in applying SF2.  
So she starts to find the vacuum state. 
This is unsuccessful for signals 8, but works for 7.  
Therefore, she can send a tuple of states with intensity $\mu_B$ between signals 3 and 7 as follows:
\begin{equation}
\begin{aligned}
	|0\rangle_1|0\rangle_1|0\rangle_2|0\rangle_2|0\rangle_3|\beta\rangle_3|\beta\rangle_4|\beta\rangle_4|0\rangle_5|\beta\rangle_5\\
	|\beta\rangle_6|0\rangle_6|0\rangle_7|0\rangle_7|0\rangle_8|0\rangle_8|0\rangle_9|0\rangle_9.
\end{aligned}
\end{equation}

Finally, in the second case Eve extracts one bit from signal 3, nothing from signal 4 since this is the control state, the bound for
the Holevo information $\chi_5=h_2[1/2 (1 - e^{-\mu_{E_1}})]$ and $\chi_6=h_2[1/2 (1 - e^{-\mu_{E_2}})]$ from signals 5 and 6, correspondingly; and nothing form signal 7. 

\begin{figure*}
	\centering
	\includegraphics[width=1\linewidth]{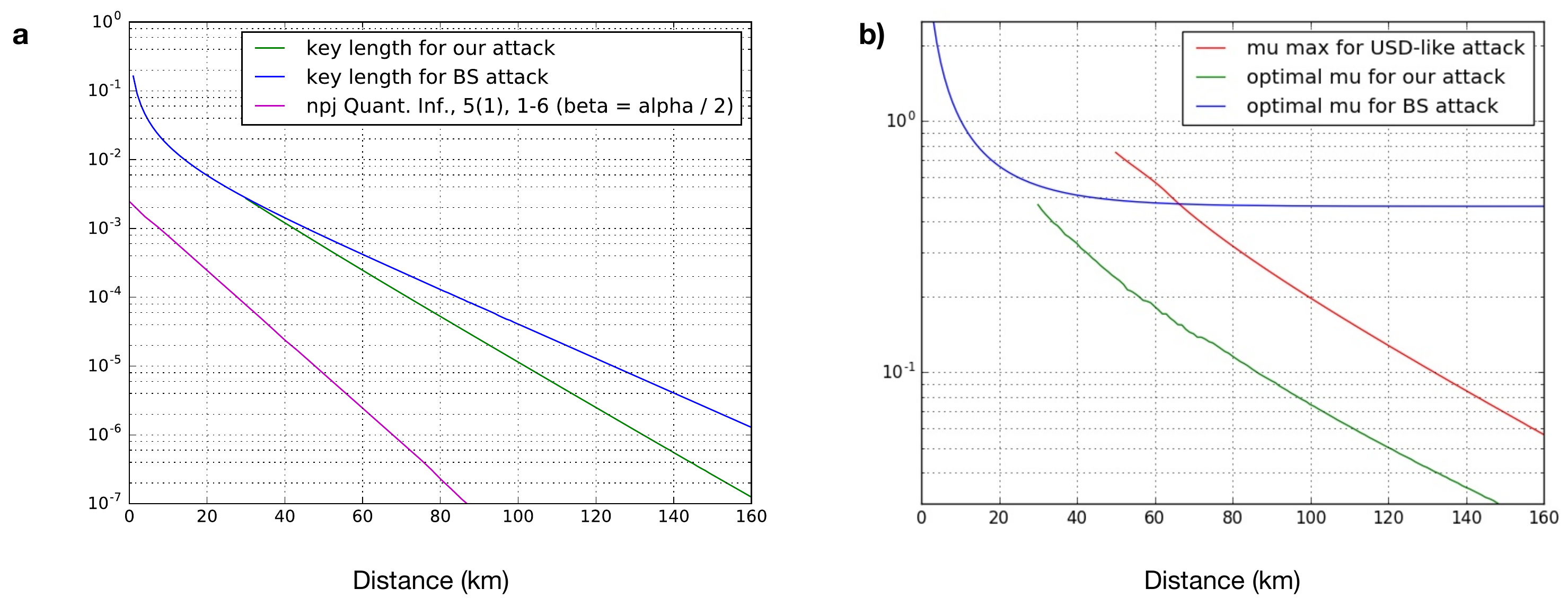}
	\vskip-4mm
	\caption{In (a) we present the key rate for proposed and beam-splitting attacks with parameters $\eta=0.1$ and $\delta=0.25$ dB/km as a function of the length of the channel between Alice and Bob.
	We also place here lower bounds for the modified version of the COW protocols given by Ref.~\cite{Lim2019} under specific conditions:
	in the curve from from Ref.~\cite{Lim2019} the different intensity of the control states is used.
	In (b) we show optimal mean photon number for three attacks (the proposed attack, BS and USD-like attacks) with parameters $\eta=0.1$ and $\delta=0.25$ dB/km as a function of the length of the channel between Alice and Bob. 
	The parameters are the same as in Ref.~\cite{Branciard2007}.}
	\label{fig:sfig1}
\end{figure*}

\subsection{Analysis}

Let us consider an expression for the successful probability of applying SF1 and SF2 operations.
The total probability is the sum of success probabilities for signal and control (decoy) states, which is as follows:
\begin{equation}
\begin{split}
	p_{SF_1}=(1-f)p_1+fq_1, \\
	p_{SF_2}=(1-f)p_2+fq_2.
\end{split}
\end{equation}
Then the fractions of the control states after successful applying SF1 and SF2 operations have the following form:
\begin{equation}
\begin{split}
	p_{d_1}=\frac{q_1f}{(1-f)p_1+fq_1}, \\
	p_{d_2}=\frac{q_2f}{(1-f)p_2+fq_2}.
\end{split}
\end{equation}

For a given tuple we can calculate the success probability and the portion of control states after forming this tuple as well as Eve's information. 
Then we can obtain an average value of the Eve's information and the control states portion as the expectation values for any given attack parameters, i.e. 
\begin{itemize}
	\item the number of SF1 operations $T_{SF1}$;
	\item the maximum number of SF2 operations $T_{SF2}$;
	\item Bob's states intensity $\mu_B$;
	\item Eve's state intensity after applying SF1 $\mu_{E_1}$;
	\item Eve's state intensity after applying SF1 $\mu_{E_2}$.
\end{itemize}
Also for a set of these parameters one can calculate the average detector click probability on the Bob side that gives the channel length between Alice and Bob, for which this attack is possible. 
The Eve's goal is to find the set of parameters maximizing her information for the given channel length with preserving the statistics of control states (if it is required). 

We note that Eve can use probabilistic strategy, where she employs various sets of parameters with corresponding probabilities.

\subsection{USD-like attack}

For the considered attack the Eve information is not equal to unity. 
If we additionally require this condition then we can construct a sort of USD-like attack, where Eve extracts full information or blocks the tuple. 
The USD-like attack corresponds to the following set of parameters: $\mu_{E_1}=\mu_{E_2}=+\infty$ and $T_{SF2}=1$. 

We note that the attack proposed in Ref.~\cite{Branciard2007} has difficulties with the neighborhood of information and control states, which can be resolved by applying our attack. 
Our USD-like attack does not block the sequence $\ket{0}\!\ket{\alpha}\!:\!\ket{\alpha}\!\ket{\alpha}$, and for it this sequence does not differ from the sequence $\ket{\alpha}\!\ket{0}\!:\!\ket{\alpha}\!\ket{\alpha}$.
We also note that our USD-like attack takes advantage of making the decision in the process of attack and is not focused on any particular strategy. 

\section{Results}\label{sec:results}

We present results of the analysis of the suggested attack in Fig.~\ref{fig:sfig1}.
Specifically, we present the key rate for our proposed attack and BS attack at the detector efficiency $\eta=0.1$ and $\delta=0.25$ dB/km as a function of the length of the channel between Alice and Bob. 
As we show in Fig.~\ref{fig:sfig1} (a) the suggested attack is better than BS attack on the whole range of channel lengths.
We note that this is not the case for the original USD attack, which is better than the BS attack starting from distances more that 100 km.
In a sense the BS attack is close to a particular case of our attack for the following set of parameters: $T_{SF2}\to\infty$, $\mu_B\approx\mu_A10^{-\delta L/10}$ and $\mu_E\approx\mu_A-\mu_B$.

In Fig.~\ref{fig:sfig1} (b) we present optimal intensities for our attack and BS attack as well as the maximum intensity starting from which Eve is able to extract full information via USD-like attack scenario. 
One can see that the suggested attack improves the results of Ref.~\cite{Branciard2007}. 
As in the case of the USD attack, there is no claim of optimality of the suggested attack based on quantum soft filtering operations.

Here we would like to mention that it is not completely correct to directly compare the key rates in Fig.~\ref{fig:sfig1}a with lower bounds from Refs.~\cite{Moroder2012,Lim2019}.
First, this is due to the fact that some the results of Refs.~\cite{Moroder2012,Lim2019} are given for the modified version of the COW QKD protocol.
The modification includes the use of the phase randomization techniques and grouping signals into blocks.
That is why we use the same protocol version as in Ref.~\cite{Branciard2007}, and we compare our results with theirs at the same parameters. 
At the same time, we plot lower bounds from Ref.~\cite{Lim2019} that are rigorously proven for any attack.
We note that we use data of Ref.~\cite{Lim2019} for the case of the detector efficiency being equal to $\eta=0.1$
(we would like to point out that the comparison between these results and Ref.~\cite{Branciard2007} that are presented in Ref.~\cite{Lim2019} are given for the different values of the detector efficiency;
moreover, some additional optimization can be provided using the techniques of Ref.~\cite{Lim2019}). 
By presenting all these results we obtain a clear picture of the security status of the COW protocol and its modifications: 
The suggested attack gives a better upper bound on the security in comparison with the results of Ref.~\cite{Branciard2007} for the {\it original} QKD protocol,
although the security can be achieved by using the modification of the protocol and the optimal parameters for the secure implementation can be obtained from Refs.~\cite{Moroder2012,Lim2019}.

\section{Conclusion and outlook}\label{sec:conclusion}

In conclusion, we have considered the applications of quantum soft filtering operations for obtaining security bounds for COW-like protocols.
We have demonstrated that in the context of the COW security analysis, quantum soft filtering operations interpolate between standard BS attack and USD technique.
We have also considered a special class of attacks based on quantum soft filtering operations, where it is required for Eve information to be unity. 
We have shown that the suggested attack outperforms the BS attack on the whole range of channel lengths and improves the attack based on the USD strategy. 
We then expect that our results are of interest for obtaining upper bounds on the security for practical QKD systems based the COW QKD protocol and its variants.

In particular, here we have considered the original version of the COW protocol. 
However, there are its modifications, which use other pairs of signals as information states. 
The phase randomization procedure (see Ref.~\cite{Moroder2012,Lim2019}) indeed could help to solve this problem. 
At the same time, we note that in principle the attack with the use of SF1 and SF2 operations still can be conducted. 
We expect that it can be harder to implement SF1 and SF2 operations, but for Alice and Bob additional difficulties in the implementation arise.
Another modification of the original COW protocol that can be considered as a countermeasure is to use decoy states of various intensity 
(these decoy states are used in a similar manner to the BB84 protocol; not to be confused with COW decoy states, since sometimes control states of the COW protocol $|\alpha\rangle|\alpha\rangle$ are also referred to as decoy states). 
In this case, soft-filtering operations should be optimized with respect to the intensity of the states, so the success probability can be quite low.
Moreover, this countermeasure requires significant changes in the original protocol.

We note that it is possible to use our approach to form a tuple of pulses which starts and ends with vacuum states (and can have vacuum states within) and therefore maintains correct visibility, but for which Eve's information is high~\cite{Moroder2012}. 
However, the precise algorithm for various modifications of the COW protocol is beyond the scope of the present work and will be considered in another place. 
We assume that this is of interest since the interesting feature of the soft-filtering operations is that they allow one to estimate the information amount in a quantum state without completly destroying its quantumness.
\smallskip

\section*{Acknowledgments}
We are grateful to N. L\"utkenhaus and N. Gisin for valuable comments, and C.C.W. Lim for useful discussion of the modified security analysis of the COW protocol. 
The research leading to these results has received funding from Russian Science Foundation under Project No. 17-71-20146.


\begin{thebibliography}{99}

\bibitem{Gisin2002}
N. Gisin, G. Ribordy, W. Tittel, and H. Zbinden,
Quantum cryptography,
{\href{https://doi.org/10.1103/RevModPhys.74.145}{Rev. Mod. Phys. {\bf 74}, 145 (2002)}}.

\bibitem{Scarani2009}
V. Scarani, H. Bechmann-Pasquinucci, N.J. Cerf, M. Dusek, N. L\"utkenhaus, and M. Peev,
The security of practical quantum key distribution,
{\href{https://doi.org/10.1103/RevModPhys.81.1301}{Rev. Mod. Phys. {\bf 81}, 1301 (2009)}}.

\bibitem{Lo2014}
H.-K. Lo, M. Curty, and K. Tamaki,
Secure quantum key distribution,
{\href{https://dx.doi.org/10.1038/nphoton.2014.149}{Nat. Photonics {\bf 8}, 595 (2014)}}.

\bibitem{Lo2016}
E. Diamanti, H.-K. Lo, and Z. Yuan, 
Practical challenges in quantum key distribution,
{\href{https://dx.doi.org/10.1038/npjqi.2016.25}{npj Quant. Inf. {\bf 2}, 16025 (2016)}}.

\bibitem{Lo1999}
H.-K. Lo and H.F. Chau,
Unconditional security of quantum key distribution over arbitrarily long distances, 
{\href{https://dx.doi.org/10.1126/science.283.5410.2050}{Science {\bf 283}, 2050 (1999)}}.

\bibitem{Shor2000}
P. Shor and J. Preskill, 
Simple proof of security of the BB84 quantum key distribution protocol,
{\href{https://dx.doi.org/10.1103/PhysRevLett.85.441}{Phys. Rev. Lett. {\bf 85}, 441 (2000)}}.

\bibitem{Biham2006}
E. Biham, M. Boyer, P.O. Boykin, T. Mor, and V. Roychowdhury,
A proof of the security of quantum key distribution,
{\href{https://dx.doi.org/10.1007/s00145-005-0011-3}{J. Cryptol. {\bf 19}, 381 (2006)}}.

\bibitem{Mayers2001}
D. Mayers,
Unconditional security in quantum cryptography,
{\href{https://dx.doi.org/10.1145/382780.382781}{J. ACM {\bf 48}, 351 (2001)}}.

\bibitem{Renner2005}
R. Renner, 
{\href{https://dx.doi.org/10.1142/S0219749908003256}{Int. J. Quantum Inform. {\bf 6}, 1 (2008)}};
{\it Security of Quantum Key Distribution},
PhD thesis, ETH Zurich; 
{\href{https://arxiv.org/abs/quant-ph/0512258}{arXiv:0512258 (2005)}}.

\bibitem{RennerGisin2005}
R. Renner, N. Gisin, and B. Kraus,
Information-theoretic security proof for quantum-key-distribution protocols,
{\href{http://dx.doi.org/10.1103/PhysRevA.72.012332}{Phys. Rev. A {\bf 72}, 012332 (2005)}}.

\bibitem{Yamamoto2003}
K. Inoue, E. Waks, and Y. Yamamoto, 
Differential phase shift quantum key distribution,
{\href{http://dx.doi.org/10.1103/PhysRevLett.89.037902}{Phys. Rev. Lett. {\bf 89}, 037902 (2002)}}.

\bibitem{Yamamoto20032}
K. Inoue, E. Waks, and Y. Yamamoto, 
Differential-phase-shift quantum key distribution using coherent light,
{\href{http://dx.doi.org/10.1103/PhysRevA.68.022317}{Phys. Rev. A {\bf 68}, 022317 (2003)}}.

\bibitem{Yamamoto2005}
H. Takesue, E. Diamanti, T. Honjo, C. Langrock, M.M. Fejer, K. Inoue, and Y. Yamamoto, 
Differential phase shift quantum key distribution experiment over 105 km fibre,
{\href{http://dx.doi.org/10.1088/1367-2630/7/1/232}{New J. Phys. {\bf 7}, 232 (2005)}}.

\bibitem{Yamamoto2006}
E. Diamanti, H. Takesue, C. Langrock, M.M. Fejer, and Y. Yamamoto, 
100 km differential phase shift quantum key distribution experiment with low jitter up-conversion detectors,
{\href{http://dx.doi.org/10.1364/OE.14.013073}{Opt. Express {\bf 14}, 13073 (2006)}}.

\bibitem{Yamamoto2007}
H. Takesue, S.W. Nam, Q. Zhang, R.H. Hadfield, T. Honjo, K. Tamaki, and Y. Yamamoto, 
Quantum key distribution over a 40-dB channel loss using superconducting single-photon detectors,
{\href{http://dx.doi.org/10.1038/nphoton.2007.75}{Nat. Photonics {\bf 1}, 343 (2007)}}.

\bibitem{Gisin2004}
N. Gisin, G. Ribordy, H. Zbinden, D. Stucki, N. Brunner, V. Scarani, 
Towards practical and fast quantum cryptography,
{\href{https://arxiv.org/abs/quant-ph/0411022}{quant-ph/0411022 (2004)}}.

\bibitem{Stucki2005}
D. Stucki, N. Brunner, N. Gisin, V. Scarani, and H. Zbinden, 
Fast and simple one-way quantum key distribution,
{\href{http://dx.doi.org/10.1063/1.2126792}{Appl. Phys. Lett. {\bf 87}, 194108 (2005)}}.

\bibitem{Stucki2009}
D. Stucki, N. Walenta, F. Vannel, R.T. Thew, N. Gisin, H. Zbinden, S. Gray, C.R. Towery, and S. Ten,
High rate, long-distance quantum key distribution over 250 km of ultra low loss fibres,
{\href{http://dx.doi.org/10.1088/1367-2630/11/7/075003}{New J. Phys. {\bf 11}, 075003 (2009)}}.

\bibitem{Stucki20092}
D. Stucki, C. Barreiro, S. Fasel, J.-D. Gautier, O. Gay, N. Gisin, R. Thew, Y. Thoma, P. Trinkler, F. Vannel, and H. Zbinden,
Continuous high speed coherent one-way quantum key distribution,
{\href{https://doi.org/10.1364/OE.17.013326}{Opt. Express {\bf 16}, 13326 (2009)}}.

\bibitem{Yamamoto20062}
E. Waks, H. Takesue, and Y. Yamamoto, 
Security of differential-phase-shift quantum key distribution against individual attacks,
{\href{https://doi.org/10.1103/PhysRevA.73.012344}{Phys. Rev. A  {\bf 73}, 012344 (2006)}}.

\bibitem{Branciard2007}
C. Branciard, N. Gisin, N. L\"utkenhaus, and V. Scarani,
Zero-error attacks and detection statistics in the coherent one-way protocol for quantum cryptography,
{\href{https://arxiv.org/abs/quant-ph/0609090}{Quant. Inf. Comput. {\bf 7}, 639 (2007)}}.

\bibitem{Branciard2008}
C. Branciard, N. Gisin, and V. Scarani,
Upper bounds for the security of two distributed-phase reference protocols of quantum cryptography,
{\href{https://doi.org/10.1088/1367-2630/10/1/013031}{New J. Phys. {\bf 10}, 013031 (2008)}}.

\bibitem{Zhao2008}
Y.-B. Zhao, C.-H. Fung, Z.-F. Han, and G.-C. Guo, 
Security proof of differential phase shift quantum key distribution in the noiseless case,
{\href{https://doi.org/10.1103/PhysRevA.78.042330}{Phys. Rev. A {\bf 78}, 042330 (2008)}}.

\bibitem{Curty2007}
M. Curty, L.L. Zhang, H.-K. Lo, and N. L\"utkenhaus,
Sequential attacks against differential-phase-shift quantum key distribution with weak coherent states,
{\href{https://arxiv.org/abs/quant-ph/0609094}{Quant. Inf. Comput. {\bf 7}, 665 (2007)}}.

\bibitem{Tsurumaru2007}
T. Tsurumaru, 
Sequential attack with intensity modulation on the differential-phase-shift quantum-key-distribution protocol,
{\href{https://doi.org/10.1103/PhysRevA.75.062319}{Phys. Rev. A {\bf 75}, 062319 (2007)}}.

\bibitem{Curty2008}
M. Curty, K. Tamaki, and T. Moroder, 
Effect of detector dead times on the security evaluation of differential-phase-shift quantum key distribution against sequential attacks,
{\href{https://doi.org/10.1103/PhysRevA.77.052321}{Phys. Rev. A {\bf 77}, 052321 (2008)}}.

\bibitem{Curty2009}
H. Gomez-Sousa and M. Curty, 
Upper bounds on the performance of differential-phase-shift quantum key distribution,
{\href{https://arxiv.org/abs/0806.0858}{Quant. Inf. Comput. {\bf 9}, 62 (2009)}}.

\bibitem{Kronberg2017}
D.A. Kronberg, E.O. Kiktenko, A.K. Fedorov, and Y.V. Kurochkin, 
Analysis of coherent quantum cryptography protocol vulnerability to an active beam-splitting attack, 
{\href{https://doi.org//10.1070/QEL16240}{Quant. Electron. {\bf 47}, 163 (2017)}}.

\bibitem{Moroder2012}
T. Moroder, M. Curty, C.C.W. Lim, L. Phuc Thinh, H. Zbinden, and N. Gisin,
Security of distributed-phase-reference quantum key distribution,
{\href{https://doi.org/10.1103/PhysRevLett.109.260501}{Phys. Rev. Lett. {\bf 109}, 260501 (2012)}}.

\bibitem{Lim2019}
Y. Wang, I.W. Primaatmaja, E. Lavie, A. Varvitsiotis, and C.C.W. Lim,
Characterising the correlations of prepare-and-measure quantum networks, 
{\href{https://doi.org/10.1038/s41534-019-0133-3}{npj Quant. Inf. {\bf 5}, 17 (2019)}}.

\bibitem{Holevo1973}
A.S. Holevo,
Bounds for the quantity of information transmitted by a quantum communication channel,
{\href{https://doi.org/10.1103/PhysRevA.78.042330}{Probl. Inf. Transm. {\bf 9}, 177 (1973)}}.

\bibitem{Curty2005}
M. Curty and N. L\"utkenhaus,
Intercept-resend attacks in the Bennett-Brassard 1984 quantum key distribution protocol with weak coherent pulses,
{\href{https://doi.org/10.1103/PhysRevA.71.062301}{Phys. Rev. A 71, 062301 (2005)}}.

\bibitem{Kronberg2014}
D.A. Kronberg,
A simple coherent attack and practical security of differential phase shift quantum cryptography,
{\href{https://doi.org/10.1088/1054-660X/24/2/025202}{Laser Phys. {\bf 24}, 025202 (2014)}}.

\end{thebibliography}
\end{document}